# Application of Flex-QA Arrays in HTS Magnet Testing

Stoyan Stoynev, Vadim V. Kashikhin, Sean Cohan, Joe DiMarco, Oliver Kiemschies, Steve Krave, Nghia Mai, Umesh Sambangi, Venkat Selvamanickam

*Abstract*— **Flexible PCB quench antennas have been very useful in providing high-quality high-resolution data in low temperature superconducting magnet tests. Similar multi-sensor arrays have been employed recently to cover a high temperature superconductor magnet tested at FNAL. In the present work, data taking conditions and magnet features to support the analysis framework are discussed. Then observations made during complete magnet powering cycles are described and analysis of quench antenna data are presented. Based on results, improvements to instrumentation and data taking are debated. Views on the future of flexible quench antenna sensors for HTS magnet diagnostics and operational support are shared.**

## I. INTRODUCTION

HIGH temperature superconductor (HTS) accelerator magnets are still in early development phase but are a major thrust in the US Magnet Development Program [1]. At present we should still expect them to operate in liquid helium environment in order to retain high Jc at high magnetic field. Thus, they feature quite high temperature margins which substantially hinders resistive voltage rise and conventional quench detection. Moreover, depending on conductor and magnet designs, voltage tap placement on the conductor, especially in large numbers, may be risky/impractical and unadvisable for work on expensive materials like HTS. For characterization of behavior, additional techniques based on acoustic sensors, fiber optics, Hall probe sensors, and others could be considered [2]. However, quench antennas (QA) may

Manuscript received xx xxx xxxx; revised xx xxx xxxx and xx xxx xxxx; accepted xx xxx xxxxu. This manuscript has been authored by Fermi Research Alliance, LLC under Contract No. DE-AC02-07CH11359 with the U.S. Department of Energy, Office of Science, Office of High Energy Physics, through the US Magnet Development Program. This work was supported by U.S. Department of Energy Office of Science, Office of High Energy Physics SBIR award DE-SC0022900. *(Corresponding author: S. Stoynev, stoyan@fnal.gov).*

S. Stoynev, V. V.. Kashikhin, S. Cohan, J. DiMarco, O. Kiemschies and S. Krave are with Fermi National Accelerator Laboratory, Batavia, IL 60510, USA.

N Mai and U. Sambangi are with AMPeers LLC, Houston, TX, 77023, USA.

V. Selvamanickam is with Department of Mechanical Engineering, Advanced Manufacturing Institute, Texas Center for Superconductivity, University of Houston, Houston, TX, 77204, USA. He has financial interest in AMPeers

Color versions of one or more of the figures in this article are available online at http://ieeexplore.ieee.org

provide the best cost-to-reliability ratio and may prove to be an invaluable type of device for HTS. A pending question is – how sensitive QA are in such settings?

In this paper we present outcomes from utilizing existing flexible quench antenna (flex-QA) arrays to cover the entire inner bore coil area of an HTS magnet model. Observed response during magnet ramps/quenches in liquid helium is summarized, based on available data from most of the ramps with quenches. Adequacy and deficiencies of the QA sensors for this application are discussed, improvements and advice for future applications are suggested. A more prominent role for quench antennas in HTS quench detection is recommended.

## II. HTS MAGNET AND IN-BORE QUENCH ANTENNA

The HTS magnet under testing was developed as part of an SBIR project with overall magnet description and previous results in liquid nitrogen reported in [3]. Detailed test results in liquid helium are reported in [4].

The magnet was a short dipole, ~50 cm of length, with aperture of 60 mm. It was made of STAR® wire [5] and COMB design and technology [6]. Fig. 1. visualizes the main geometry of the magnet which was tested in liquid helium within the FNAL's VMTF [7] in early 2024. The magnet test featured warm bore tube with in-bore support for flex-QA arrays [8]. A flex-QA panel with slanted channel orientation (2 x 10 channels) and another one with 10 straight channels featuring bucked signals were employed, both described geometrically and conceptually in [8], [9], Fig. 2. The two panels were placed on top of each other, well aligned, and wrapped around the QA support, Fig. 2. Each channel had a width of 8 mm. Each of the panels covered ~ 345 degrees of azimuthal angle with the "missing" 15 degrees centered around one of the magnet poles. A 3D Hall probe, which was embedded inside the tubular carbon fiber support and was well aligned to it and the QA, helped with proper positioning of the QA with respect to the magnetic field and magnet. The QA azimuthal orientation was known within 5 degrees and longitudinal position – within 3 mm. Uncertainties came from misalignment of the Hall probe axes with respect to the support, difficulties keeping proper orientation when QA is repositioned (the Hall probe was away from the QA in axial direction), determination of the magnetic center. Radially, the QA positions were well known, certainly within 1 mm, and were almost 14 mm away from the closest conductor in layer 1, Fig. 1; this distance is far from desirable.

Once overlayed, the two flex-QA arrays had their channels crossing (on a surface at a fixed diameter) which allowed localization of observed electro-magnetic events [9]. The readout rate of 100 kHz and precision of 16-bit along with continuous data taking during whole current ramps gave us the opportunity to analyze response behavior at both magnet quenching and during magnet ramping.

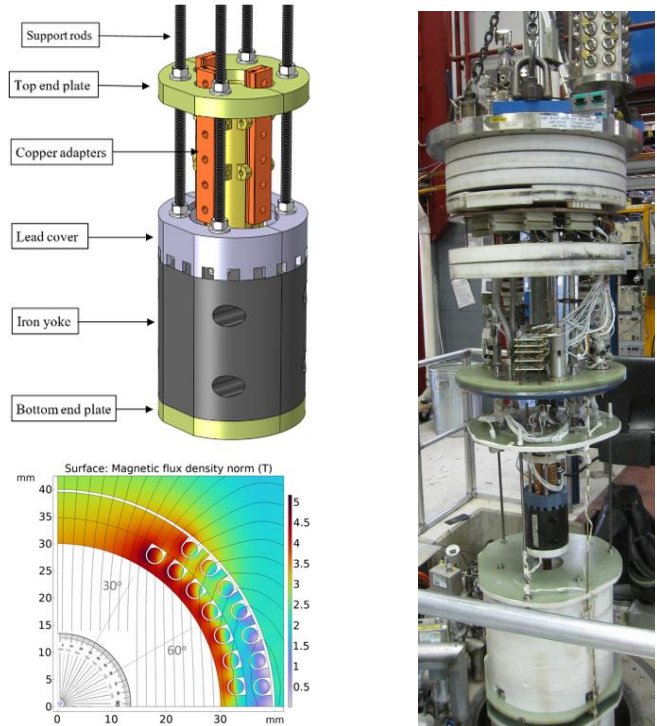

**Fig. 1.** COMB-STAR-1 magnet – design (top), and magnetic field distribution at design current, along with bore geometry (bottom); the magnet in preparation for testing (right).

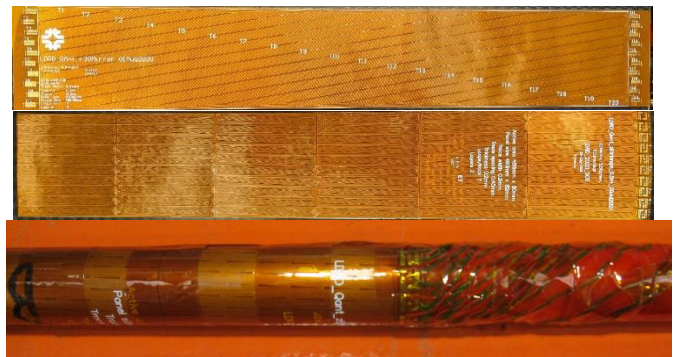

Fig. 2. Top two: Flex-QA used in magnet testing – two types of QA are employed; Bottom: quench antenna panels wrapped precisely around their support (270 mm diameter, visualized on Fig.1).

## III. QUENCH ANTENNA RESPONSE

A flex-QA array is not the only QA type to characterize quenches [10], [11], [12] and typically requires a large number of channels. However, it is a reliable and simple device, and we were particularly interested in sensitivity to HTS "disturbances" for purposes of future quench characterization and detection (with reduced channel count). One of the main advantages of the flex-

QA arrays is their ability to fit to existing coil surfaces and minimize distance to the conductor, thus maximizing flux through sensor windings. Our pre-existing setup did not allow to fully explore this feature. Besides, the few voltage taps employed did not allow to separate quenches between the two layers in the magnet. We are planning to address those deficiencies in future HTS magnet tests, opting to rely on simplicity and a step-by-step approach in the first one.

The recorded quench events were in many aspects similar, in terms of quench antenna response. We concentrate on one of the ramps to quench (#3), with the highest quench current. We assess it is a fair qualitative representation of the other ramps as well, no major differences need reporting for this analysis.

### A. Quench Initiation

Multiple current ramps ended with quenches, at different ramp rates and temperatures of liquid helium, 1.9 K or 4-4.5 K [4]. All of them were spontaneous quenches. QA data were taken for eight quenches at each 1.9 K and higher temperatures of liquid helium. QA signals, approximately consistent with the time of voltage rise observed (with limited time resolution due to noise), were seen for the first six of the recorded quenches at 1.9 K and nothing was seen above background for the rest of the quenches. Importantly, the signals observed had little-to-none temporal evolution ("shape"), Fig. 3, which is very different from observations in low temperature superconducting magnets (Rutherford cables) [13], [14], [15], [9]. With the present setup we are unable to distinguish quenching in the two different layers and we cannot rule out that the inner layer acts as a screen to the outer one. This may explain why we do not detect any signal for some quenches. Alternatively, the QA sensitivity in our non-optimized setup is too low for those quenches, especially at 4-4.5 K where processes were shown to be slower [4]. Later setups will be modified to resolve those ambiguities. It should be noted that in LTS no screening effect was observed with flex-QA [9]. Regardless, signals, when present, and their relative amplitudes clearly followed patters allowing to identify quench location by crossing channels, like [9]. That aspect of analysis is less relevant in the context on this paper.

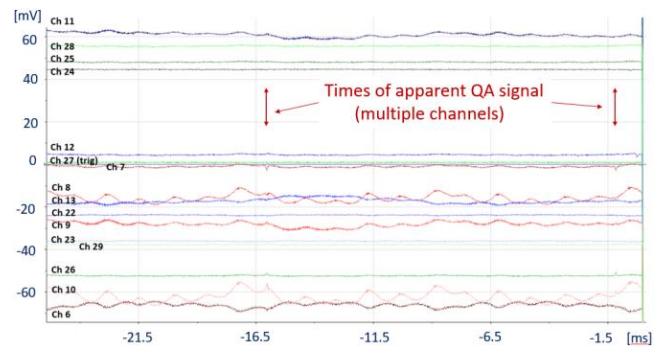

Fig. 3. QA response before quench detection in one of the current ramps. Channels with "signal" and some neighboring channels are shown. Channel Ch27 is the trigger channel – quench detection flips its state to higher voltage (5 V) at 0 ms. Voltage offsets for channels are random and have no meaning.

Fig.3 shows that channels in the "bucked" QA (all channels above #19) deliver stable readout with minimal noise, as intended by design. The remaining channels, from the "slanted" QA, have larger repeatable and in principle reducible noise. Channel enumeration follows geometrical order. It is instructive to see that the "bucked" channels' signal amplitude, at ~ -16 ms as marked, is smaller and signal itself more localized – in fact only channel Ch26 clearly responds above noise with channel Ch29 a distant exception. This is partially due to the QA design, with the edges of channels mostly insensitive to current variations [9]. On the other hand, multiple slanted QA channels respond to the quench (associated current redistribution), all of them - adjacent channels; the sign of the observed peak (at ~ -16 ms) changes from channel #8 (and #7) to channel #9 (and #10), while the maximum amplitude is observed at channels #7 and #10. As explained in [9], sensors of this QA type are insensitive to current alternations along the channel mid-width; the reason they are slanted with respect to the conductor axis is to avoid quench propagation along the sensor axis. This means that current redistribution (flux change) originating immediately under a sensor will yield no or little response in that sensor while adjacent channels on both sides will pick it up, with different signs! What we observe in practice are two mid-channels with small signal and two adjacent channels with large(r) signal. Those indicate the source of the "disturbance" (quench) is located somewhere in the area of channels #8 and #9, where the signal sign flips. The signal is well aligned in time between channels and is seen at ~ 16 ms before quench detection which is based on coil voltage readout. Later spikes are seen in some of the QA channels within 2 ms of quench detection. If we were to rely on QA for detecting this quench we would have gained those 16 ms for magnet protection, and not worry about how slow coil-voltage rise was or if we could promptly register it above noise before it is too late [4]. We are to argue why this may be a plausible strategy for detection.

According to earlier tests [9] the DAQ system response to sudden current change features a relaxation time of ~0.2 ms. The "spikes" observed in the present work have similar relaxation (decay) times. Given the small characteristic time of those signals, it is important to account for the system response. Then, it is apparent that what we observe is likely localized and nearly instant current redistribution in the STAR® wire.

### B. Current Ramps

The "bucked" channels show remarkable stability during current ramping for hundreds of seconds - there are very few clear "events" (spikes) above background for any channel, Fig. 4. This can be compared to the "slanted" channels which not only appear as noisier, although most of the noise is reducible, but also more "eventful". It is worth noticing that the noise-band of both types of channels widens in the middle of the first 100 s range which corresponds to several hundred of Amperes (10 A/s initial ramp rate). The effect is real and may relate to flux jumps, a standard occurrence in LTS-based magnets, but needs to be studied separately. Data reveals the ability of the "bucked" QA to completely suppress some events ("global") and be sensitive to others ("local"). It is an important feature to explore for quench detection.

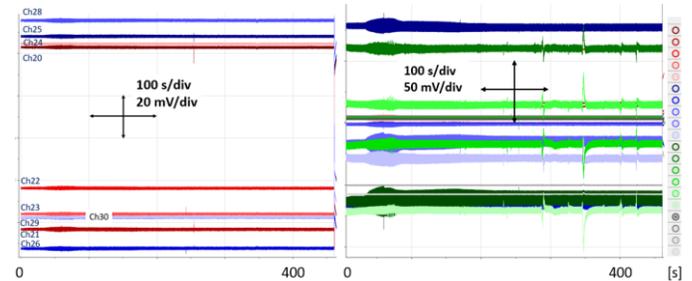

Fig. 4. QA response during the reviewed current ramp: "bucked" channels (left) and "slanted" channels (right). Some of the latter only cover non-conducting coil elements (the color scheme is in geometrical order). "Zero" time is the beginning of data taking, just before ramping starts.

### IV. SIGNAL CHARACTERISTICS IN BUCKED CHANNELS

It is difficult to make the quench detection case for the "slanted" channels without additional data processing/filtering. It is however undisputable that "bucked" channels offer much sturdier performance, as long as sensitivity is not in question. Thus, we will discuss the "bucked" type of QA sensors. Fig. 5 presents more information on the channel responding to the quench, Ch26, for the typical ramp we are detailing. The peak-to-peak noise in the "bucked" channels is consistently 1-2 mV, except in a narrow range in the beginning (around event E1). Noise across all channels is further reducible – harmonic power supply noise (60-720 Hz) is the dominant contributor, in "slanted" channels as well. Noise filtering is yet to be explored.

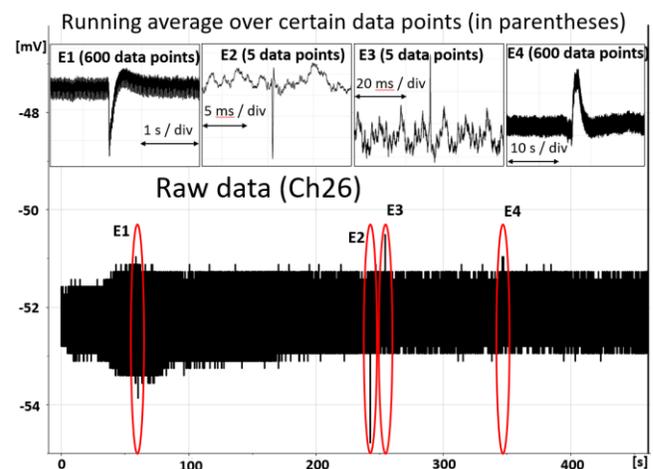

Fig. 5. Detailed view of the "bucked" channel Ch26 response during ramping and observable "events" before the quench. Running average of 600 data points reveals the timescale of long "events", while 5 data points smooths some of the random noise. Those help visualizing event characteristics.

As Fig. 5 shows, without filtering there are only four events visible, i.e. with amplitude beyond noise level, before the quench event which is presented separately on Fig. 6.

As we discussed, channel Ch26 is the one among the bucked channels showing visible above-noise response at quench with signal-to-noise ratio of 5, Fig. 6 (the "intrinsic"/random noise is much less than 1-2 mV, as observed). Channel Ch29 is the only other one with discernable signal, Fig. 3, with signal-to-noise ratio of 3. The quench signals observed in all quench antennas (in all events with detected quenches) are spike-like, given the instrumentation characteristic response (decay) time of ~0.2 ms [9]. This also applies for events E2 and E3 on Fig. 5, which are not quench events and which are visually distinguishable from the mostly harmonic noise. However, both of those events are near simultaneously observed in multiple quench antenna channels, eventually with different signal shapes, Fig. 7.

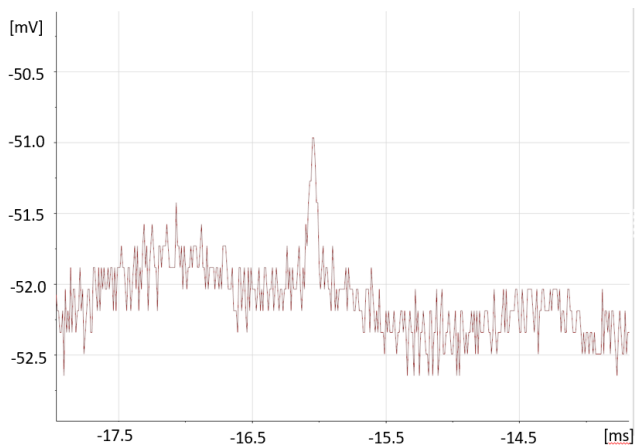

Fig. 6. Response to the quench in channel Ch26, with respect to quench detection time (for the ramp of interest).

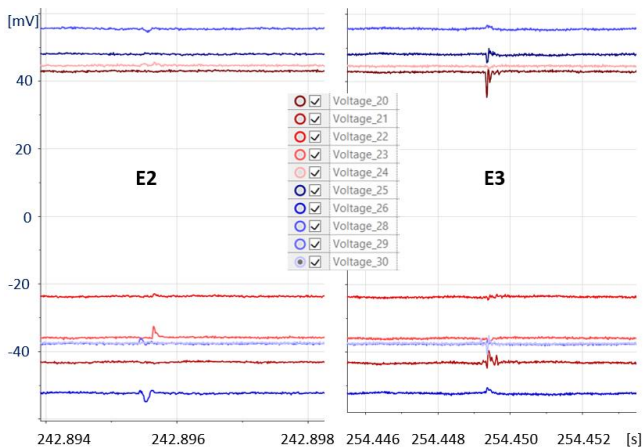

Fig. 7. All "bucked" QA signals in two "events" (following timing in Fig. 5) – E2 (left) and E3 (right).

Similarly, the non-spike-like events E1 and E4 on Fig. 5 are also observed throughout the QA coverage; besides, their characteristic time is orders of magnitude larger than E2 and E3 (and the quench event), Fig 5. In this sense, none of those pre-quench events is "local" while the quench event is "local".

## V. DISCUSSION

The way the quench antenna array was employed pertained to its use as a simplified but universal quench characterization option. The signal is reversely proportional to distance squared and the fact that the flex-QA responds to quenches at such a sub-optimal distance, Fig. 2 and Fig. 1 (bottom) suggests the following:

- without modifications flex-QA designs of the type employed can be adapted to fit on coil surfaces to reliably detect quench initiation events in HTS; the distance to the closest super-conductor can be as low as 1 mm with corresponding gain in sensitivity

- the existing QA trace density is more than enough in terms of sensitivity; it could be increased if necessary

- one could explore smaller "bucking" regions to improve on locality or noise reduction, trading for lower but sufficient sensitivity

- one could "OR" channels or numerically encode them – something being developed at FNAL – to reduce number of readout channels

Given that "bucked" channels are much cleaner in terms of "raw"/real-time noise, it is conceivable to think that quench detection would benefit from relying on those sensors rather than "non-bucked" ones. It is still unclear if QA sensors are shielded by a coil layer in which case each conductor layer needs to face its own flex-QA. It is also possible to have quenches with no response in the QA. We cannot claim our data support any of those, nor they exclude them. We will experiment with those possibilities in the next HTS magnet tests where more appropriate instrumentation will be in place.

The strength and thickness (~ 0.2 mm) of the QA panels allows for applications within the structure of the magnet and for multilayer arrays. Those could be used as backups for detection channels. It is practical to have crossed sensors forming a grid and relying on it for characterization. A combination of them could be processed independently for quench detection purposes, possibly with channel encoding.

The goal of our experiments is to accumulate data from COMB with round REBCO conductor magnet models, compare to wire/cable measurements and create simulation models that can reasonably describe observations. Those will serve as a solid ground not only for further studies but will provide reliable support for quench detection development based on flex-QA. So far observations strongly back up the possibility to realize such a system for COMB magnet designs.

## V. CONCLUSION

Flex-QA arrays were deployed in the warm-bore of the COMB-STAR-1 magnet model. Despite inferior geometrical configuration, the sensors demonstrated remarkable sensitivity to quenches and the "bucked" version of the QA showed quite manageable noise levels during magnet powering. Questions remain regarding "invisibility" of some of the quenches, reasonable explanations are to be explored in future tests. Data so far suggest there is a viable path to develop a quench detection system for HTS magnets based on flex-QA.